%% file: paper.tex
\documentclass[10pt,conference,compsocconf,letterpaper]{IEEEtran}

\usepackage{makeidx}
\usepackage[usenames,dvipsnames,svgnames,table]{xcolor}
\usepackage{xspace}
\usepackage{amsmath}
\usepackage{amssymb}
\usepackage{mathtools}
\usepackage{mathptmx}
\usepackage[font={small,it}]{caption}
\usepackage{tikz}
\usepackage{tikz-qtree}
\usepackage{graphicx}
\usepackage{multirow}
\usepackage{booktabs}
\usepackage{tablefootnote}
\usepackage{amsmath}
\usepackage{flushend}
\usepackage[all]{nowidow}

\usepackage{url}

\usepackage{paralist}

\usepackage{enumitem}

\usepackage{cleveref}
\newcounter{centredequ}
\newenvironment{centredequ}{\refstepcounter{equation}\hfill\begin{math}}{\end{math}\hfill$(\theequation)$\par\noindent}
    \crefname{centredequ}{eq.}{eqs. }
    \Crefname{centredequ}{Eq.}{Eqs. }

\newcommand{\name}{RITM\xspace}

\newcommand{\myparagraph}[1]{
    \noindent
    {\bf #1.}
}

\begin{document}

\title{\name: Revocation in the Middle}

\author{
    \IEEEauthorblockN{Pawel Szalachowski, Laurent Chuat, Taeho Lee, and Adrian Perrig}
    \IEEEauthorblockA{
    Department of Computer Science\\
    ETH Zurich, Switzerland
    }
}

\maketitle

\begin{abstract}
Although TLS is used on a daily basis by many critical applications, the
public-key infrastructure that it relies on still lacks an adequate revocation
mechanism. An ideal revocation mechanism should be inexpensive, efficient, secure, and
privacy-preserving. Moreover, rising trends in pervasive encryption pose new
scalability challenges that a modern revocation system should address. In this
paper, we investigate how network nodes can deliver certificate-validity
information to clients. We present \name, a framework in which middleboxes (as
opposed to clients, servers, or certification authorities) store
revocation-related data. \name provides a secure revocation-checking mechanism
that preserves user privacy. We also propose to take advantage of
content-delivery networks (CDNs) and argue that they would constitute a fast and
cost-effective way to disseminate revocations. Additionally, \name keeps
certification authorities accountable for the revocations that they have issued,
and it minimizes overhead at clients and servers, as they have to neither store
nor download any messages. We also describe feasible deployment models and
present an evaluation of \name to demonstrate its feasibility and benefits in a
real-world deployment.
\end{abstract}

\section{Introduction}
\label{sec:intro}
\input{intro}

\section{Background and Related Work}
\label{sec:pre}
\input{pre}

\section{\name}
\label{sec:details}
\input{details}

\section{Deployment Models}
\label{sec:deploy}
\input{deploy}

\section{Security Analysis}
\label{sec:analysis}
\input{analysis}

\section{Implementation}
\label{sec:implementation}
\input{implementation}

\section{Evaluation}
\label{sec:eval}
\input{eval}

\section{Discussion and Practical Considerations}
\label{sec:discussion}
\input{discussion}

\section{Conclusion}
\label{sec:conclusions}
\input{conclusions}

\bibliographystyle{abbrv}
\bibliography{references}
\end{document}

%% file: intro.tex
Checking the revocation status of a certificate is a fundamental process for the
security of any connection established via a public-key infrastructure (PKI).
Certificate revocation occurs when a non-expired certificate must be
invalidated. In order to revoke a certificate, the certification authority (CA)
that issued the certificate adds its \textit{serial number}\footnotemark~to a
\textit{blacklist}. Then, the CA tries to publish or disseminate the most
recent version of that list. This process is sometimes referred to as the Grand
Challenge of PKIs~\cite{gutmann2004build}, for several
reasons:
\begin{inparaenum}[\itshape 1)]
  \item To keep connections secure, revocations must be accessible and the
    dissemination should be fast, so that every interested party is able to deny
    an illegitimate connection (that relies on a revoked certificate) as soon as
    possible.
  \item The revocation system should be robust and resilient to server
    misconfiguration and blocking attacks.
  \item The revocation status should be authentic (i.e., all clients can verify
    that any given revocation was created by an authorized CA).
  \item The system should preserve user privacy (i.e., no third party, such as a
    CA, should be able to learn about the servers that clients tried to contact).
\end{inparaenum}
\footnotetext{The serial number is a positive integer value (represented at most
by 20 bytes), assigned uniquely to every CA-issued certificate~\cite{rfc5280}.}

Several revocation schemes have been proposed and deployed over the last few
years, but the current situation is still unsatisfactory in many aspects. For
instance, CRLs and OCSP violate user privacy and are inefficient, while the
deployment of OCSP stapling requires server reconfiguration. Due to these
drawbacks, browser vendors recently decided to push special revocation lists to
clients using software update mechanisms. We consider that such a scheme is not
future-proof, as even now the revocation lists issued by these vendors are
limited to a small fraction of all certificates (for efficiency reasons), and
the number of certificates/revocations is expected to grow.

Revocation systems will face new scalability challenges in the near future, due
to the increasing security awareness of website owners and hosting providers,
and new initiatives that aim to make TLS ubiquitous.  For instance, the
\textit{Let's Encrypt}~\cite{letsencrypt} project provides a free certificate to
any website administrator who requests it, as well as a convenient
certificate-management software. The project, during few months of operation,
issued more than one million certificates which protect approximately 2.4
million domains~\cite{letsencrypt}. The \textit{Universal SSL}
project~\cite{universal_ssl} initiated by CloudFlare constitutes another
prominent example of mass TLS adoption. It provides over 2 million clients with
free TLS certificates. Furthermore, hardware-limited/mobile devices and
paradigms such as the Internet of Things (IoT) may rapidly increase the number
of TLS clients. As a consequence, we may expect that the number of TLS
connections will increase by orders of magnitude, in the near future.
Unfortunately, mass TLS adoption poses scalability problems for current
revocation systems, as pervasive TLS deployment would result in an increased
number of revocations. As reported by Liu et al.~\cite{liu-2015-revocation},
even today, due to efficiency reasons, all mobile browsers have disabled
revocation checks.

Catastrophic events, such as the disclosure of the infamous Heartbleed
vulnerability~\cite{Zhang:2014:ASC:2663716.2663758,durumeric2014matter}, may
again result in a huge number of revocations. In the current setting, revocation
propagation, besides operational issues, reportedly induces significant
financial costs for CAs~\cite{hidden_cost}. This might lead to a situation in
which mass-revocation decisions are strongly influenced by financial concerns.

In this paper, we present \name, the first scheme to incorporate revocation
functionalities into middleboxes. The main intuition behind our solution is that
clients do not have to store any revocation lists nor need to establish any new
connection (to check revocation status) if all of their TLS traffic goes through
a middlebox that can provide them with the correct revocation status. \name
consists of two mechanisms: a dissemination network that provides revocation
messages to middleboxes, and a protocol that allows middleboxes on a TLS
client-server path to deliver revocation status to clients.

\name minimizes overhead at clients and servers, and can be easily implemented
with low operational costs as part of an efficient and profitable deployment
model. The system is resilient against a range of well-known attacks against the
current revocation schemes (like blocking or server reconfiguration). Moreover,
following the transparency trend of the PKI ecosystem, \name keeps CAs
accountable for the revocations they issue, and it minimizes the probability of
a successful attack in which a CA would maliciously alter a revocation. We
evaluate \name by showing its performance, overhead, cost-effectiveness, and
feasibility in a real-world deployment. We also demonstrate that \name can
disseminate a new revocation within seconds to all supported parties across the
Internet.

%% file: pre.tex
\myparagraph{Hash Chains and Hash Trees}\label{sec:pre:hash_trees}
We denote a cryptographic hash function as $H(.)$. A \textit{hash chain} of
length $m$ is the successive application of a hash function $H(x)$ ($m$ times)
to an input $x$. The $m$-th value of a hash chain is denoted as $H^m(x)$, for
example $H^3(x) = H(H(H(x)))$ and $H^0(x) = x$.  Using hash chains it is
possible to compute $H^m(x)$ from any $H^i(x)$ if $i<m$, but it is infeasible to
compute $H^m(x)$ from any value $H^j(x)$ where $j>m$. In a \textit{hash tree}
every non-leaf node is a hash of its child nodes. This structure allows to
efficiently (logarithmically in the number of leaves) prove that a given node is
an element of the tree. When the leaves are sorted in a predefined order, it is
also possible to efficiently prove that a given element is absent.

\myparagraph{Content-Delivery Network}
A CDN is a distributed infrastructure dedicated to efficiently deliver content
to clients. CDNs are mostly driven by web content, and their main goal is to
reduce latency and the load exerted on \textit{distribution points}. To operate
efficiently, a CDN consists of many \textit{edge servers}, geographically
distributed in the proximity of large collections of clients. Edge servers
replicate the original content by obtaining it from a distribution point.  The
dominant communication paradigm used in CDNs is \textit{data pulling}, i.e.,
clients pull the content from edge servers, which similarly pull it from the
distribution points. To work efficiently, edge servers cache the content for a
specific \emph{time to live} (TTL), which usually is set by the origin. From the
client's perspective, the CDN is an abstraction, as routes to edge servers are
not made explicit.

\myparagraph{Desired Properties}
The revocation system must be able to handle \textit{near-instantaneous
revocations}, i.e., revocation information should be accessible by the
interested parties within seconds. This property is crucial to minimize the
\textit{attack window}, which, with the current revocation systems, varies from
hours to days.  A revocation system must have the ability to meet this
requirement in order to support applications such as e-banking, online trading,
or transaction processing from point-of-sale terminals.  We stress that this
property implies that a client should also be notified about a revocation in the
middle of established connections.  Otherwise, a \textit{race condition} is
possible when the client establishes a connection just before the corresponding
certificate is revoked.  This is especially important for applications that
establish long-lived TLS connections, like VPNs, TLS tunnels, remote
filesystems, secure messaging, or environments like IoT (where devices are
reluctant to conduct costly TLS handshakes).

The revocation process must not require \emph{server reconfiguration/update}, or
availability of an \emph{additional third party}. Ideally, whenever a CA has
initiated the dissemination of a revocation message, and a server is accessible,
then the clients that connect to this server should be able to access the
message. Whenever the effectiveness of a scheme depends on the configuration of
the server, an adversary who compromised the server can decrease the
effectiveness of the scheme by tuning configuration parameters.

Revocation information must be \textit{authentic}. Only the legitimate party
(i.e., a CA that issued a certificate) can create a fresh revocation message,
and everyone must be able to verify that message. Moreover, certificate
revocation status should be obtained without sacrificing \textit{privacy}.
Especially, it should not be required that clients contact a third party, as it
reveals which domains they contacted.

The system must be \textit{efficient} in terms of transmission, computation, and
storage. It should not introduce any significant latency to a client-server
connection. The solution must \textit{scale} with an increasing number of
revocations, certificates, servers, CAs, and clients. The revocation system must
also provide or fit into \textit{feasible and profitable deployment} models.

Following current trends in the field of PKIs, CA actions related to revocation
issuance must be \textit{transparent} and \textit{accountable}. Anyone must be
able to verify that the appropriate CA issued a given revocation, and the system
must provide means to ensure that everyone has the same view of revocations.
This guarantees that a whole range of potential attacks conducted by a malicious
CA will be detected.

\myparagraph{Related Work and its Drawbacks}\label{sec:related}
The first attempt to address the revocation problem, proposed to distribute the
information through \emph{Certificate Revocation Lists} (CRLs)~\cite{rfc5280},
published by CAs at CRL distribution points (specified in the certificate).  To
verify a certificate's validity, a client downloads a CRL and checks whether the
certificate is listed in the CRL. Unfortunately, this approach  is inefficient
as an entire CRL must be downloaded to verify a single certificate, and some
CRLs are currently large (a few megabytes).  As this download is performed
during a TLS handshake, it significantly increases the latency of the entire
process.  \emph{Delta CRLs} can reduce this large overhead by allowing clients
to only fetch new revocations when required, but they still suffer from other
drawbacks of regular CRLs.  In particular, a CA can violate user privacy by
creating a dedicated distribution point for a targeted certificate, and simply
determine which client contacts a targeted domain. Furthermore, Gruschka et al.
report that during a 3-month period, only 86.1\% of the CRL distribution points
were available~\cite{gruschka2014analysis}.

The \emph{Online Certificate Status Protocol} (OCSP)~\cite{santesson2013x} was
proposed to address the inefficiency of CRLs. With OCSP, a client can contact a
CA to obtain the current revocation status of a single, given certificate. This
solution, however, is still inefficient because the CA may be under heavy load,
and an extra connection during the TLS handshake is still required. Furthermore,
OCSP has a serious privacy issue, as the CA learns which server the client
connects to.

To solve these problems \emph{OCSP Stapling} was
proposed~\cite{pettersen2013transport}. With this solution, the server
periodically fetches an authenticated OCSP response from the CA, and then sends
the response stapled along with the certificate in subsequent TLS connections.
Unfortunately, this solution requires servers to update their software. Although
a study suggests that some of the major domains (i.e., 22.5\% of domains that
are sampled from the Alexa's list of most popular sites) have deployed OCSP
stapling~\cite{sslpulse2015}, the overall deployment rate is still marginal---a
study showed that only 3\% of all certificates are served by servers that
adopted OCSP stapling~\cite{liu-2015-revocation}. Moreover, the age of the
stapled response can be customized by a configuration parameter; therefore, a
long attack window can be introduced by an adversary or a
misconfiguration~\cite{stark2012case}.

\emph{Short-Lived Certificates} (SLCs)~\cite{rivest1998can,topalovic2012towards}
solve problems associated with CRLs and OCSP by eliminating revocation
completely. SLCs are designed to be valid for a few days, which leaves a long
attack window as SLCs are irrevocable for that time period. In addition, their
deployment depends on software update and configuration on the server side, as
every server must be set up to contact a CA periodically.

Recently, browser vendors decided to disseminate special CRLs
(CRLSet~\cite{langley2012revocation}, OneCRL~\cite{mozilla_rev}) through
software updates and major browsers use these lists as the main revocation
mechanism. Such an approach does not require any server update or
reconfiguration; however, a long attack window still exists, as software updates
are infrequent and as clients apply them at irregular points in time (with a
heavy-tail distribution). This technique also faces scalability issues. Every
client must store a complete list, which may be infeasible or undesired for
hardware-limited devices~\cite{liu-2015-revocation}. As the dissemination is
done through unicast communication, scalability in the face of an increasing
number of clients is challenging. Finally, with an increasing number of
revocations, CRLSets may become too large to be delivered to all clients. To
address this issue, CRLSets include only a marginal number of revocations
(0.35\%, as reported~\cite{liu-2015-revocation}). Such a policy restricts the
method's scalability and effectiveness. For this reason, the CLRSet approach has
been criticized~\cite{chrome_crl,liu-2015-revocation}.

\emph{RevCast}~\cite{schulman2014revcast} is another recent approach, which
improves revocation dissemination through unique properties of radio broadcast.
RevCast proposes an architecture in which CAs broadcast revocation messages, and
clients with radio receivers can receive and collect them immediately. RevCast
is still a CRL-like approach where clients must possess the entire CRL. To
satisfy this requirement, an additional infrastructure must be created (e.g., a
device must obtain and store new CRL entries when clients are not listening to
the broadcast transmission). RevCast also requires that users purchase and
install radio receivers. Another downside of RevCast is that it requires to
significantly change CA operations, and the maximum bandwidth is 421.8 bit/s,
which prevents handling high revocation rates rapidly.

Log-based approaches~\cite{laurie2012revocation, KimHuaPerJacGli13,
Basin:2014:AAR:2660267.2660298, PKISN} make the revocation issuance process
transparent and accountable, as CAs are obligated to submit revocations to
public and verifiable logs. Unfortunately, the deployment of these schemes is
mainly server-driven (i.e., servers have to be configured to fetch and serve
fresh revocation status) or client-driven (i.e., clients contact logs, which
compromises their privacy). Moreover, the attack window is large, as logs are
designed to update their internal state every few hours.

\myparagraph{System Model}
We investigate the revocation problem for a TLS communication in which only the
server is authenticated to the client. Such a setting is typical on today's
Internet. We use TLS as our default environment, but we stress that our scheme
can be combined with any protocol that exchanges certificate(s) in
plaintext.\footnote{Protocols that encrypt handshakes (like the draft version of
TLS 1.3~\cite{rescorla2016transport}), would need to expose a server's
certificate or its identifier in plaintext.} We assume that
\begin{inparaenum}[\itshape 1)]
\item CAs can reach a dedicated dissemination network at regular
    intervals~(denoted $\Delta$)\footnote{The value of $\Delta$ is a trade-off
    between the size of the attack window and efficiency. Throughout the paper
    we analyze values from 10 seconds to 1 day.  For the sake of simplicity, we
    assume that $\Delta$ is a global parameter, but this assumption can be
    easily relaxed---see \S\ref{sec:discussion}.},
\item TLS and the cryptographic primitives that we use are secure, and
\item the different parties are loosely time synchronized.
\end{inparaenum}
Time is expressed in Unix seconds and the $time()$ function returns the current
time. We denote a message $msg$ signed by $X$'s private key $K_{X}^{-}$ as
$\{msg\}_{K_{X}^{-}}$.

\myparagraph{Adversary Model}
We assume that an adversary can control the network (can modify, block, and
create any message) and can compromise any of its elements. The adversary's goal
is to convince non-compromised clients either that a revoked certificate is
still valid or that a valid certificate is revoked.

%% file: details.tex
\myparagraph{Lessons Learned and Insights}
We summarize some of the most important lessons learned from the deployment of
previous revocation schemes (for more details see \S\ref{sec:related}):
\begin{inparaenum}[\itshape 1)]
	\item \textit{During the connection establishment process a client cannot make
		any dedicated connection to check whether a certificate is revoked.}
    Otherwise, an additional latency is introduced, privacy may be violated, and
    the connection can be established only when the third party is available.
    (See CRL and OCSP.)
	\item \textit{Due to large bandwidth and storage overheads, clients cannot be
	  constantly equipped with complete lists of revocations.} Moreover, due to
    the expected increase in the number of TLS clients and servers, this might
    become even more problematic in the future. (See CRL and CRLSet.)
	\item \textit{It is very unlikely that a significant fraction of all servers
    be rapidly updated to use a new security enhancement.} (See OCSP Stapling
    and log-based approaches.)
\end{inparaenum}

Our first observation is that, as clients and servers cannot be updated with
fresh revocations and clients should avoid obtaining revocation status via a
dedicated connection, the only remaining possibility to provide revocation
status to clients is to implement a network functionality that will support it.

The second observation is that an efficient, robust, and fast dissemination
network is a necessary component of a satisfactory revocation system in the
current TLS ecosystem, and CDN infrastructures fulfill all these requirements.

Finally, the current formats of revocation lists do not allow to efficiently
monitor the revocations issued by CAs. For instance, it is challenging to detect
a misbehavior when a CA shows different CRLs (or OCSP messages) to different
clients (for instance to hide the fact that a given certificate is revoked).

\myparagraph{High-level Design}
Taking into consideration the lessons learned and our observations, the
high-level idea is to
\begin{inparaenum}[\itshape 1)]
  \item propagate revocations through a CDN,
  \item piggyback revocation status to standard TLS communications, and
  \item enhance the format of revocation lists to enable a more efficient
    monitoring of CAs.
\end{inparaenum}
In this setting, every revocation is distributed through a dissemination
network, and for every TLS connection, a dedicated middlebox (that is connected
to the dissemination network) on the path between the client and the server will
put fresh, authentic, and accountable revocation status along with the native
TLS message towards the client.

\begin{figure}[h!]
	\centering
    \includegraphics[width=0.8\linewidth]{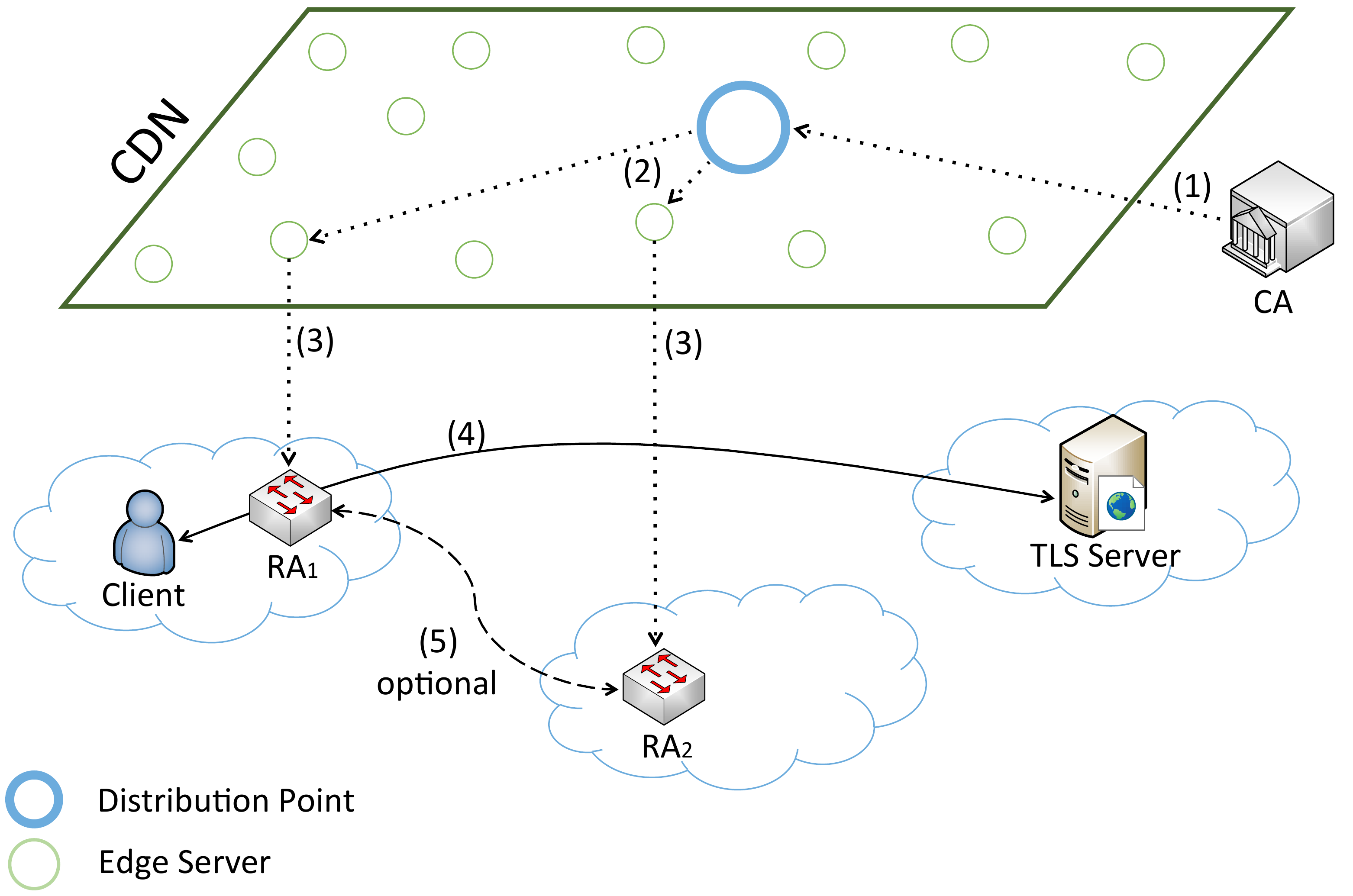}
    \caption{High-level architecture of \name.}\label{fig:diss_net}
\end{figure}

\name's high-level architecture is depicted in Fig.~\ref{fig:diss_net}. The
essential element of \name is a network middlebox called the \textit{Revocation
Agent} (RA). RAs are designed to fulfill several purposes. First, they connect
to the dissemination network used to propagate revocations. As opposed to client
machines, middleboxes are usually always turned on, which allows them to
constantly receive and collect new revocations from the dissemination network
(steps 1, 2, 3). Then, for every TLS connection, an RA provides fresh revocation
status to the \name-supported clients (step 4). For non-supported clients and
for non-TLS traffic, RAs act as transparent middleboxes. Finally, RAs are able
to keep CAs accountable, as they monitor the revocations they issue, compare
their views of revocations with other parties of the system, and report any
detected misbehavior in a provable manner (step 5).

\myparagraph{Revocation Lists}\label{sec:details:dict}
One challenge is to find a format of revocation list that meets our
requirements.  Ideally, it should be realized in a way that allows to keep CAs
accountable and that supports efficient, authentic, and fresh proofs of
certificate status.  To achieve this, \name extends the concept of
\textit{authenticated dictionary}~\cite{naor2000certificate}. This structure
allows a non-trusted \textit{prover} to prove to a \textit{verifier} that the
queried object is (or is not) an element of a given dictionary created by a
trusted third party. In our scenario, the RA is a prover, the client acts as a
verifier, and the CA is a trusted third party. Every CA maintains a dictionary
of its own revocations, while every RA stores copies of all the dictionaries.
These dictionaries are constantly updated (by the CAs through the dissemination
network) with new revocations and metadata, which ensures that a given
dictionary is fresh and consistent. Dictionaries are append-only
(\S\ref{sec:discussion} discusses this property and optimizations), and the
proof generated by a prover:
\begin{inparaenum}[\itshape 1)]
    \item can efficiently express presence/absence of a given revocation in/from
      a given dictionary,
    \item is authentic, i.e., the verifier is sure that the statement was
      produced by the CA, and
    \item is timestamped, so that a verifier can reject a stale proof.
\end{inparaenum}
Moreover, authenticated dictionaries allow to keep CAs accountable, as they help
to maintain a consistent view across the system. For instance, entities with a
short proof can check whether their dictionary views are consistent.
Consequently, all inconsistencies can be eventually detected and proven.

\begin{figure}
	\centering
	\footnotesize
	\setlength{\tabcolsep}{10pt}
	\noindent\rule{\linewidth}{0.5pt}

	\begin{description}
		\item[\texttt{insert}] (executed by a CA that revokes a certificate with
      serial number $s_x$):
	\end{description}

	$n$ : size of the dictionary with $s_x$ appended, \quad $m$ : parameter chosen
  by CA

	\begin{enumerate}
	  \item Insert $s_x,n$ into the tree and rebuild it.
	  \item For random value $v$ compute $H^m(v)$.
	  \item Return \textbf{signed root}:
	    \begin{centredequ}\label{eq:signed_root}
	      \{\textit{root}, n, H^m(v), \textit{time}()\}_{K_{\textit{CA}}^{-}}
	    \end{centredequ}
	\end{enumerate}

	\begin{description}
		\item[\texttt{update}] (executed by an RA on input
      $s_x, \{\textit{root}, n, H^m(v), t\}_{K_{\textit{CA}}^{-}}$):
	\end{description}

	\begin{enumerate}
	  \item Verify the signature and timestamp $t$.
	  \item Insert $s_x,n$ into the tree and rebuild it.
	  \item Accept the applied changes only if the newly-built root equals
      $\textit{root}$ and $n$ is a number of leaves in the new tree.
	\end{enumerate}

	\begin{description}
		\item[\texttt{refresh}] (executed by a CA at least every $\Delta$ if there
      is no new revocation):
	\end{description}

	\begin{description}
		\item $t$ : timestamp of the latest signed root
	\end{description}

	\begin{enumerate}
	    \item $p = \lfloor(\textit{time}() - t)/\Delta\rfloor$.
	    \item If $p < m$ return \textbf{freshness statement}:
	        \begin{centredequ}\label{eq:fresh_statement}
	            H^{m-p}(v).
	        \end{centredequ}
	    \item If $p \geq m$ create a new signed root (see
	        Eq.~(\ref{eq:signed_root})) and go to step 1.
	\end{enumerate}

	\begin{description}
		\item[\texttt{prove}] (executed by an RA on input $s_x$):
	\end{description}

	\begin{enumerate}
	  \item Produce presence/absence proof for $s_x$.
	  \item Return \textbf{revocation status}:
	    \begin{centredequ}\label{eq:proof_root}
	      \textit{proof}, \{\textit{root}, n, H^m(v), t\}_{K_{CA}^{-}}, H^{m-p}(v),
	    \end{centredequ}
	    where $H^{m-p}(v)$ is the latest (current) freshness statement.
	\end{enumerate}

	\noindent\rule{\linewidth}{0.5pt}
	\caption{Interactions with an authenticated dictionary. Operations
	\texttt{insert} and \texttt{update} can be performed in batch (to add multiple
	revocations to the tree simultaneously).}\label{fig:auth_dict_api}
\end{figure}

\name implements authenticated dictionaries with hash trees (see
\S\ref{sec:pre:hash_trees}).  Every leaf of a tree is a serial number of the
revoked certificate concatenated with the number of that revocation.  For each
dictionary, revocations are numbered consecutively, starting from~1.  The
numbering ensures that revocations are inserted into the tree in the correct
order, which helps to maintain a consistent copy of the dictionary and detect
attacks (such as revocation reordering). Leaves are sorted in lexicographical
order (using their serial numbers), which allows to prove that a given leaf is
or is not an element of the tree. We define the message names and the
dictionary's interface in Fig.~\ref{fig:auth_dict_api}.

\myparagraph{Dissemination}\label{sec:details:diss}
\name can employ an existing CDN infrastructure as a dissemination network, and
such a deployment brings many benefits. CDNs are efficient and robust in
large-scale deployments~\cite{al2011overclocking}, and we show that they provide
an excellent environment for dissemination of revocation messages. As our study
shows, a CDN-based deployment is also cost-effective (see
\S\ref{sec:eval:cost}).

\begin{table}[b!]
\centering
\footnotesize
\begin{tabular}{@{}lcc@{}}
	\toprule
    \textbf{Time} & \textbf{Revoked serial number} & \textbf{Disseminated message} \\
    \midrule
    $t = t_0$ & $s_a$, $s_b$, $s_c$ & $s_a$, $s_b$, $s_c$, $\{\textit{root}, n, H^m(v), t\}_{K_{\textit{CA}}^{-}}$ \\
    $t = t_0 + \Delta$ & none & $H^{m-1}(v)$ \\[1ex]
    $t = t_0 + 2\Delta$ & none & $H^{m-2}(v)$ \\[1ex]
    $t = t_0 + 3\Delta$ & $s_d$ & $s_d$, $\{\textit{root}', n+1, H^m(v'), t\}_{K_{\textit{CA}}^{-}}$ \\
    \bottomrule
\end{tabular}
\caption{Example of messages disseminated over time.}\label{tab:diss}
\end{table}

Whenever a CA wishes to revoke some certificate(s), it updates its local
dictionary through the \texttt{insert} operation (see
Fig.~\ref{fig:auth_dict_api}). Then, the CA contacts the network's distribution
point with the \textit{revocation issuance} message, i.e., the revoked serial
number(s) with a new signed root (e.g., at times $t_0$ and $t_0 + 3\Delta$ in
Tab.~\ref{tab:diss}). The distribution point verifies this message and initiates
the dissemination process by sending the message to edge servers. Eventually,
all RAs obtain the message from the corresponding edge server and update their
local copy of the CA's dictionary through the \texttt{update} operation (see
\S\ref{sec:details:dict}).

\begin{figure}
    \centering
    \includegraphics[width=0.9\linewidth]{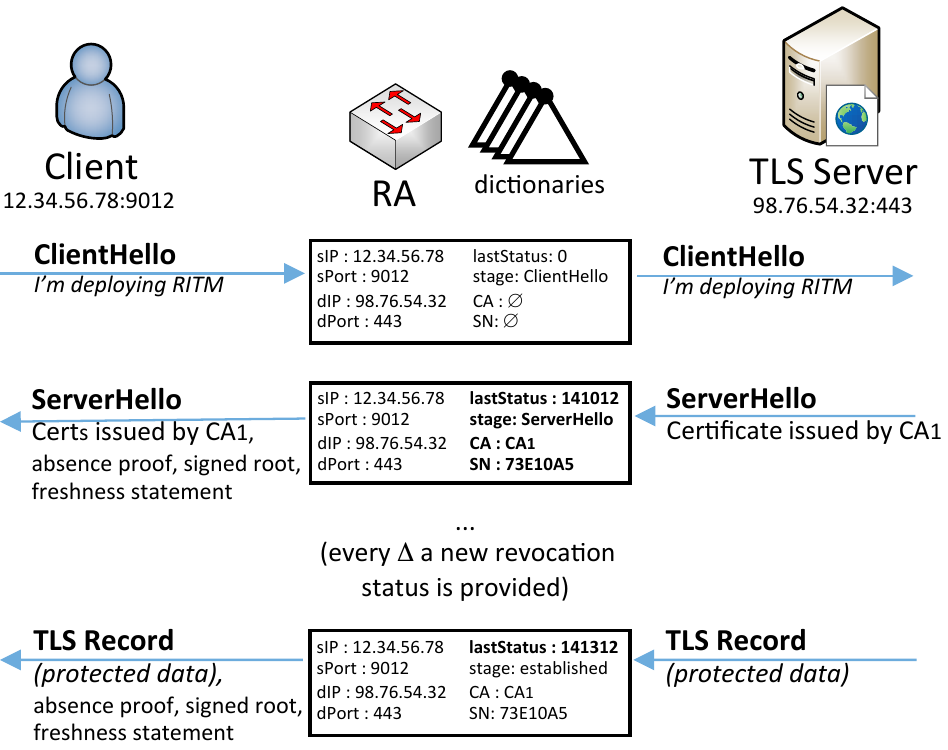}
    \caption{An example of \name-supported TLS connection.}\label{fig:example}
\end{figure}

When no new revocations are issued during a time period $\Delta$, CAs are still
obliged to keep their dictionaries fresh. In order to achieve this, every CA
creates and disseminates a freshness statement (Eq.~\ref{eq:fresh_statement})
periodically (e.g., at times $t_0 + \Delta$ and $t_0 + 2\Delta$ in
Tab.~\ref{tab:diss}). Therefore, to keep dictionaries fresh when there is no new
revocation within a time period $\Delta$ (which is a common case), CAs
disseminate only a freshness statement that in practice is significantly shorter
than a signed message.  Moreover, due to hash chain properties
(see~\S\ref{sec:pre:hash_trees}), the message is authentic, as only a CA can
create a valid freshness statement for a given time period. The dissemination
process of a freshness statement is the same as before, and RAs, after
verification, replace the freshness statement with the new one.

\name provides a simple synchronization protocol to keep the dictionary copies
of RAs correctly updated.  As every revocation issuance message contains a value
$n$ that denotes the number of revocations that the CA has issued, an RA can
easily detect whether his local copy is desynchronized from CA's original copy
based on $n$. Whenever a desynchronization is detected, the RA contacts an edge
server specifying the number of valid consecutive revocations it has observed.
The signed root with a timestamp (Eq.~(\ref{eq:signed_root})) also guarantees
integrity, as it is impossible to alter a dictionary without changing the root.

\myparagraph{Validation}\label{sec:details:valid}
Our technique relies on the fact that the negotiation phase of TLS is performed
in plaintext. Thus, RAs can detect new TLS connections and the corresponding
server certificates that must be validated. A \name-supported TLS connection is
presented in Fig.~\ref{fig:example}, and we describe below the different steps
in detail.

\begin{enumerate}[leftmargin=*]
  \item The TLS connection is initiated by a \texttt{ClientHello} message. The
    client sends it with a dedicated TLS extension~\cite{rfc5246}, to inform
    potential RAs on the path that this TLS connection should be protected by
    \name.

  \item An RA on the path observes the traffic for TLS messages. Whenever it
    receives a \texttt{ClientHello} message with the \name extension, it creates
    the following state:
    \begin{equation}\label{eq:state0}
      \begin{gathered}
      	\footnotesize
        \texttt{sIP=}\textit{IP}_c\texttt{, dIP=}\textit{IP}_s
        \texttt{, sPort=}\textit{port}_c\texttt{, dPort=}\textit{port}_s\texttt{,} \\
        \footnotesize
        \texttt{lastStatus=0, stage=ClientHello, CA=$\emptyset$, SN=$\emptyset$,}
      \end{gathered}      
    \end{equation}
    \texttt{sIP, dIP, sPort, dPort} denote source/destination IP addresses and ports;
    \texttt{lastStatus} is the latest time at which a revocation status was sent
    to the client; \texttt{stage} describes the current stage of the TLS
    connection; \texttt{CA} allows the RA to identify the correct dictionary;
    and \texttt{SN} is the certificate's serial number. The RA identifies
    supported TLS connections using this state information.

    When the state is created, the \texttt{ClientHello} message is passed to
    the next hop on the path towards the server.

  \item The server responds with a \texttt{ServerHello} message followed by
    the server's certificate. (The server ignores the \name's
    \texttt{ClientHello} extension.)

  \item When an RA receives a \texttt{ServerHello} message that matches the
    previously-created state, it inspects the content of the message to
    obtain the server's certificate and to determine the CA that issued the
    certificate. Then, the RA executes the \texttt{prove} procedure (see
    Fig.~\ref{fig:auth_dict_api}) over the CA's dictionary. A revocation status
    (i.e., proof, signed root, and a freshness statement as in
    Eq.~(\ref{eq:proof_root})) is appended to the \texttt{ServerHello} message
    and passed to the client. The RA updates the state from
    Eq.~(\ref{eq:state0}) by setting
    \texttt{lastStatus} to the current time, \texttt{stage} to
    \texttt{ServerHello}, \texttt{CA} to the CA identifier, and
    \texttt{SN} to the serial number of the server's certificate.

  \item When the client receives a \texttt{ServerHello} message with a
    revocation status (Eq.~(\ref{eq:proof_root})), he removes the status from
    the message (to not influence the TLS protocol), and a standard
    certificate-validation procedure is executed.  Additionally, the client
    verifies the revocation status sent by the RA. The server's certificate
    is accepted when:
    \begin{enumerate}
        \item it passes the standard validation,
        \item the revocation status contains a valid revocation absence proof
        (validated against a signed root), and
    	\item the freshness statement is no older than $2\Delta$ (see
        \S\ref{sec:analysis:short}), i.e., for received revocation status
        as in Eq.~(\ref{eq:proof_root}), $H^{p'}(H^{m-p}(v))$ or
        $H^{p'+1}(H^{m-p}(v))$ equals $H^{m}(v)$, where
        $p' = \lfloor(\textit{time}() - t)/\Delta\rfloor$.
    \end{enumerate}

  \item Finally, when the TLS connection is accepted (i.e., when the server
    sends a \texttt{Finished} message), the RA updates the \texttt{stage} field
    to \texttt{established}.  From this moment on, the client and the server
    communicate securely. Whenever the RA detects that $\Delta$ time has passed
    since it has received a revocation status for a supported connection (i.e.,
    $\textit{time}() - \texttt{lastStatus} \ge \Delta$), the RA uses the first
    TLS packet from the server to the client to piggyback a fresh revocation
    status (similarly as in step 4). After the fresh status is sent, the RA
    updates \texttt{lastStatus} to $\textit{time}()$.

  \item Whenever the client receives the packet with a revocation status, he
    checks the status, similarly as in step 5.
\end{enumerate}
Steps 6 and 7 are conducted periodically, at least every $\Delta$, and the
connection is interrupted by the client, when a fresh absence proof is not
provided. Whenever a supported connection is finished or timed out, the RA
removes the corresponding state.

\name supports two mechanisms of TLS resumption, namely \textit{session
identifiers}~\cite{rfc5246}, and \textit{session tickets}~\cite{rfc5077}.
Although the TLS handshake for these modes is abbreviated, the presented
mechanism is similar. We present \name in a setting where the revocation status
is provided to the client piggybacked on standard TLS traffic. However, in
\S\ref{sec:discussion} we discuss other implementation choices to achieve this
functionality.

\myparagraph{Consistency Checking}\label{sec:details:consist}
\name enables RAs and clients to monitor CA actions related to certificate
revocations. Through the construction of trusted dictionaries and validation
logic, a CA has to provide freshness statements, which are short, unique, and
unforgeable pieces of information about a dictionary's content. This message can
be used directly by RAs to monitor the consistency of dictionaries. The goal is
to ensure that a given party of the system has the same view of the dictionaries
as the rest of the system.  In order to achieve this property, an RA can
periodically request a random edge server for its copy of the signed root. Only
by comparing local and downloaded values, the RA can confirm that a copy stored
by the edge server is identical to the RA's copy. Simply exchanging the latest
signed root is enough to keep CAs accountable, as dictionaries are append-only,
and as a violation from the append-only property is easily detectable.

Due to nature of the CDN environment an RA can detect and contact only a limited
number of edge servers (usually, a list of the closest edge servers can be
obtained via a DNS query). This limitation may decrease the effectiveness of the
consistency checking procedure. To address this problem, a \textit{map server}
can be introduced. That server would store addresses of RAs (and optionally edge
servers), so that they can communicate directly to exchange their current
freshness statements. An alternative way is to deploy a gossip protocol (for
example, as proposed by Chuat et al.~\cite{gossip2015}), where clients would
exchange signed roots obtained during TLS connections, or to deploy dedicated
parties to monitor CAs as proposed in log-based
approaches~\cite{KimHuaPerJacGli13}.

Such procedures can protect from various possible attacks launched by a
malicious CA (for details see \S\ref{sec:analysis}). Moreover, any detected
misbehavior can be cryptographically proved and reported, for example to the
software vendors.

%% file: deploy.tex
Incremental deployment is a challenging process for almost all new security
technologies. The main challenges, associated with the deployment of \name, are
the following:
\begin{inparaenum}[\itshape 1)]
    \item How can the \name-supported clients be protected from downgrade
      attacks (in which the attacker convinces the client that there is no RA on
      a path towards the server)?
    \item How can the new system be deployed in a backward-compatible manner?
    \item How can \name be efficiently deployed, i.e., so that a small number
    	of RAs support all TLS traffic?
\end{inparaenum}

We present two deployment models for \name, which we believe are feasible with
current networks and that address the above-mentioned problems. The intuition
behind the presented models is that \name can work effectively when RAs are
placed close to clients or servers, as it is easier to mitigate downgrade
attacks and ensure that all TLS traffic passes through the RA. In
\S\ref{sec:discussion} we discuss other deployment strategies and aspects.

\myparagraph{Close to the servers}\label{sec:deploy:dc}
Our first deployment model assumes that RAs are placed close to the servers,
e.g., at the ingress point of a data center or a server farm. Typically, a
load-balancer that distributes workload across multiple servers by inspecting
the content of the packet is placed at the ingress point of a data center. For
encrypted traffic (via TLS), the load-balancer additionally functions as a
\textit{TLS terminator}, which establishes and maintains TLS connections on
behalf of the servers~\cite{delignat2015network}, to terminate TLS connections
and inspect the decrypted traffic. RA functionality can be augmented to TLS
terminators so that every TLS connection between a data center and its clients
could benefit \name. In this scenario, the augmented TLS terminators put an
indication that the server side supports \name within an extensions of TLS'
\texttt{ServerHello}. This approach eliminates downgrade attacks on TLS
connections since an adversary cannot undetectably alter the TLS handshake
messages.

\myparagraph{Close to the clients}\label{sec:deploy:lan} Alternatively, RAs can
be placed close to the clients, e.g., close to the gateways of clients' access
networks. In fact, an RA can even be combined with the gateway, minimizing
setup and deployment costs for the network operator that operates clients'
access network. This approach provides two benefits: it protects client's
connections to any TLS server regardless of deployment at the servers, and the
network operator can place RAs in his network in such a way that all traffic in
the network will traverse through a RA, hence, offering protection to the
entire network. To this end, RAs can be either placed at the \textit{choke
points} of the network, or SDN-based solutions can be used~\cite{Qazi2013}.

However, the client-side deployment comes with a security vulnerability: it is
susceptible to downgrade attack by an adversary that is within the client's
access network and tunnels client's traffic to an external network so that the
traffic bypasses a RA in the network.  To avoid such downgrade attack, a
network operator must provide clients with an authentic information that his
network supports \name. This can be realized by secure network management
protocols~\cite{Droms:2001:ADM:RFC3118}.

%% file: analysis.tex
\myparagraph{Short Attack Window}\label{sec:analysis:short} Through a fast and
robust dissemination and with an appropriate certificate-acceptance policy
(i.e., clients may accept a certificate only with its fresh revocation status),
\name reduces the long attack windows associated to many revocation schemes.
Effectively, the attack window is 2$\Delta$.  Although CAs publish and RAs
download updates every $\Delta$, they do not have to be synchronized. Due to the
communication technique of CDNs (i.e., pull), a situation where an RA downloads
an update just before the CA publishes new changes may occur.  Hence, a
\textit{tolerance} parameter is required, and $\Delta$ is set as a value of this
parameter.

The value of the $\Delta$ parameter is a trade-off between security and
efficiency. However, as an RA is an online device connected to the dissemination
network, we may expect even low values (tens of seconds) as feasible for a
production setting (see \S\ref{sec:eval:speed}).  Revocation checking in \name
is independent from server configuration, hence it is impossible to introduce a
long attack window through a server-configuration parameter.

\myparagraph{Race Condition}
For an established TLS connection, client and RA perform periodic revocation
checks.  Such a design protects from a race condition in which a long-lived TLS
connection with a server is established just before the corresponding
certificate is revoked. In such a case, the client would unintentionally
communicate with a non-trusted party until the connection is terminated. To the
best of our knowledge, \name is the first revocation system that protects from
such an attack.

\myparagraph{MITM and Blocking Attack}
A MITM adversary, without the ability to bypass an RA cannot get any advantage.
As it is assumed that the adversary cannot break the TLS protocol, he cannot
undetectably modify the \texttt{ClientHello} message which carries \name's
extension. This message informs the RA on the path that the connection should be
supported by \name.  Such an adversary can drop or delay status messages carried
on TLS traffic but, according to \name's validation policy (see
\S\ref{sec:details:valid}), this would lead to a connection interruption.

\myparagraph{Downgrade Attack}
An adversary able to tunnel TLS traffic, in order to avoid an on-path RA, can
try to launch a downgrading attack (as outlined in \S\ref{sec:deploy:lan}),
convincing a client that there is no RA on the path towards the server. This
attack is generic and powerful against the majority of security protocols that
are partially deployed.

To eliminate such an attack, the client must be provided with authentic
information that the given connection is supported by an RA. Protection from
this attack depends on the deployment scenario and we provide two feasible
solutions.  In the first deployment scenario  \name is deployed by TLS
terminators (see \S\ref{sec:deploy:dc}). In this case, the client would receive
a deployment confirmation from a TLS terminator, within a \texttt{ServerHello}
message. This message is itself protected by the TLS protocol, thus the
adversary cannot forge it. As a consequence, the adversary is not able to launch
a successful downgrade attack against \name. Second, \name is deployed in a
network, which can provide clients authentic information about \name support.
For instance, this information can be delivered through a secure equivalent of
the DHCP protocol.  After the bootstrapping, the client would expect a \name
protection for every TLS connection, so that even an adversary able to tunnel
the traffic cannot bypass the scheme.

We do not discuss downgrade attacks against the TLS protocol itself, such as
HTTPS stripping attacks~\cite{marlinspike2009more}, as this goes beyond the
scope of this paper. However, we stress that this problem has been investigated
in the past and some effective countermeasures are currently
deployed~\cite{rfc6797,rfc7469}.

\myparagraph{RA and Dissemination Network Compromise}
Another advantage of \name is that it requires a small trust base (i.e., only
CAs are trusted).  \name is thus resilient to RA compromises. As the
construction of authenticated dictionaries allows RA and CDN to be untrusted, an
adversary that compromises an RA and CDN cannot get any benefit. Trusted
dictionaries are created by CAs and, to forge a revocation status~(see
Fig.~\ref{fig:auth_dict_api}), the adversary would need to forge a digital
signature, or break the hash function. Both conditions lead to the
contradiction, as it is assumed that cryptographic primitives are secure.
Moreover, an adversary with control over RAs cannot suppress any revocation
message, as clients expect to see fresh revocation status messages periodically.

\myparagraph{Misbehaving CA}
Furthermore, \name makes detectable a whole range of attacks that may be
conducted by a misbehaving CA. The dictionaries are append-only, authenticated
structures, and a signed root (Eq.~\ref{eq:signed_root}) uniquely corresponds to
a particular version of the dictionary. Therefore, it is possible to detect that
a CA is misbehaving, for instance if the CA shows one version of the dictionary
to part of the system, and another version to the rest. To prove that, it is
enough to find two different signed roots with the same dictionary size. This
can be realized in many ways (see \S\ref{sec:details:consist}). For example, RAs
can randomly contact CDN edge servers or other RAs and  compare their
locally-stored statements with the newly-downloaded ones. It is sufficient to
compare only the latest signed  roots (with the same dictionary size), as
dictionaries are append-only; therefore, whenever a misbehaving CA creates two
versions of a dictionary, then the CA must constantly maintain these two
different dictionaries (to satisfy update procedure from
\S\ref{sec:details:dict}). Such a procedure guarantees that the system has a
consistent view of all issued revocations. The append-only property and
consecutive revocation numbers also protect from attacks such as revocation
reordering, or revocation deletion. This substantially limits the possibilities
of a misbehaving CA.

\name can be also extended to the scenario where a compromised CA wishes to
revoke its own certificate  without introducing a collateral damage (i.e.,
without revoking certificates issued before the compromise). To handle such a
case, a method presented by Szalachowski et al.~\cite{PKISN} can be adopted.

\myparagraph{More powerful adversaries}
The attacks in which an adversary can capture a CA and another element of the
system are especially dangerous. For instance, when an RA and a CA are
compromised at the same time, then the adversary can create a fake version of
the CA's dictionary (e.g., without a given revocation), and provide a proof to
any client, using this fake dictionary. Although this scenario is unlikely,
\name can be extended to make even such an attack visible. As clients receive
proofs accompanied with signed roots, they could contact edge servers, other
RAs, or even other \name-supported clients, to ensure that their view of the
dictionary is consistent. The exact details of such a detection mechanism go
beyond the scope of our current work, but our system can be for instance
enhanced by gossip protocols~\cite{gossip2015}.

\myparagraph{Privacy}
\name preserves user privacy. Clients do not need to establish any extra
connection for checking the status of a certificate and only the entities on the
path towards the server know which clients contact which servers. In particular,
CAs and the CDN network are not able to obtain this information.

%% file: implementation.tex
\label{sec:impl:impl}
To prove feasibility of \name's deployment we implemented it and the
implementation includes
\begin{inparaenum}[\itshape 1)]
	\item authenticated dictionaries,
	\item an RA communication module (to contact the dissemination network and
	  update dictionaries),
  \item a deep packet inspection (DPI) module for RAs to analyze packets,
    identify new TLS connections, and keep track of the connection state, and
	\item an RITM-supported client.
\end{inparaenum}
\name is implemented in Python (2.7.6), and we used Scapy with the Scapy-SSL/TLS
module to realize DPI of TLS connections. We used the SHA-256 hash function, but
we truncated its output to the first 20 bytes.  Also, to optimize the bandwidth
and computational overhead, we used the Ed25519~\cite{bernstein2012high}
signature scheme, which allows to obtain a digital signature of 64 bytes.

RAs are implemented as userspace daemons. All network traffic goes through an RA
that verifies whether a packet belongs to the TLS handshake protocol or a
previously established (and active) \name-enabled connection. Then, the RA
follows the different steps described in \S\ref{sec:details:valid}.  For every
TLS connection (after receiving a \texttt{ServerHello} message) \name-supported
TLS client expects a revocation status from the corresponding CA (delivered by
an RA), which states that the certificate is not revoked.  After this check
passes, the client continues the standard TLS validation.  For the communication
between an RA and the CDN, we built a simple HTTP(S)-based API.  Every $\Delta$,
each RA contacts an edge server via an \texttt{HTTP GET} request to pull new
revocations and freshness statements. Our implementation of RAs can be adapted
to operate with any CDN infrastructure.

%% file: eval.tex
\subsection{Dataset used}\label{sec:eval:dataset}
\begin{figure}
  \centering
  \includegraphics[width=0.75\linewidth]{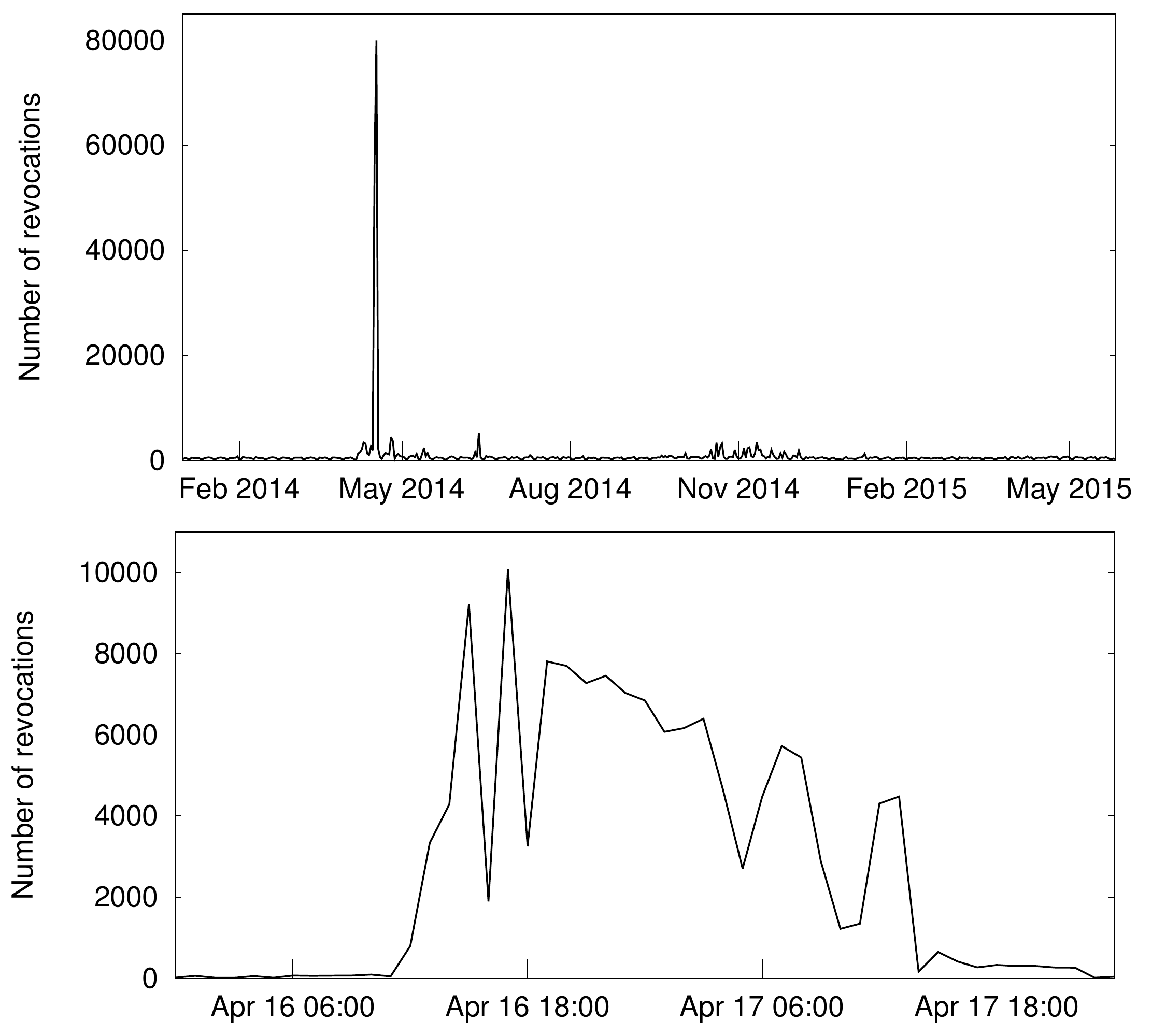}
  \vspace{-2pt}
  \caption{Number of revocations issued between January 2014 and June 2015 with
  a focus on the Heartbleed peak.}
  \label{fig:heartbleed}
\end{figure}

We use a dataset collected by the \textit{Internet Storm
Center}~\cite{isc_crl}, which contains the most complete list of CRLs (to the
best of our knowledge), namely, 254 separate revocation lists. From these
lists, we identify 1,381,992 unique revocations and an average of 5,440
revocations per CRL. Moreover, we identify that 3 bytes are used most frequently
(32\% of all revocations) as the size of serial numbers. Thus, we use 3-byte
serials numbers throughout this analysis.

To show how \name reacts to extremely high revocation rates, we investigate
revocations issued in April 2014. This time period is especially interesting
because it includes the disclosure of the infamous
\textit{Heartbleed}~\cite{hb_link} OpenSSL vulnerability.  This bug allowed
remote attackers to read protected server memory, including private keys
associated with certificates. Consequently,  thousands of certificates were
revoked over a few days. The detailed analysis of this event is presented by
Durumeric et al.~\cite{durumeric2014matter} and Zhang et
al.~\cite{Zhang:2014:ASC:2663716.2663758}.  Fig.~\ref{fig:heartbleed} (top
graph) represents the number of  revocations observed from January 2014. The
Heartbleed event is visible as a sudden peak in the middle of April 2014. The
highest revocation rate was observed on 16--17 April 2014 (bottom graph).

\subsection{Speed of Dissemination}\label{sec:eval:speed}

\name strives to make revocations as readily accessible as
possible, and one key criteria towards this goal is to minimize the time
needed by RAs to download revocation messages. Using Amazon
CloudFront~\cite{amz_link} as an example
CDN, we measure the time required for an RA to download
revocation messages. Amongst many different CDN networks, we chose Amazon
CloudFront because it allows to turn off content caching
at the edge servers (by setting the caching interval to zero, i.e., TTL=0). This
feature allows us to measure the worst-case scenario performance (in terms of
latency), since the content
needs to be fetched from the origin server for every request.
Furthermore, since RAs in different geographical areas may experience different
download latency, we take measurements from various locations. To
this end, we use the Planetlab~\cite{chun2003planetlab}.

\begin{figure}
  \centering
    \includegraphics[width=0.8\linewidth]{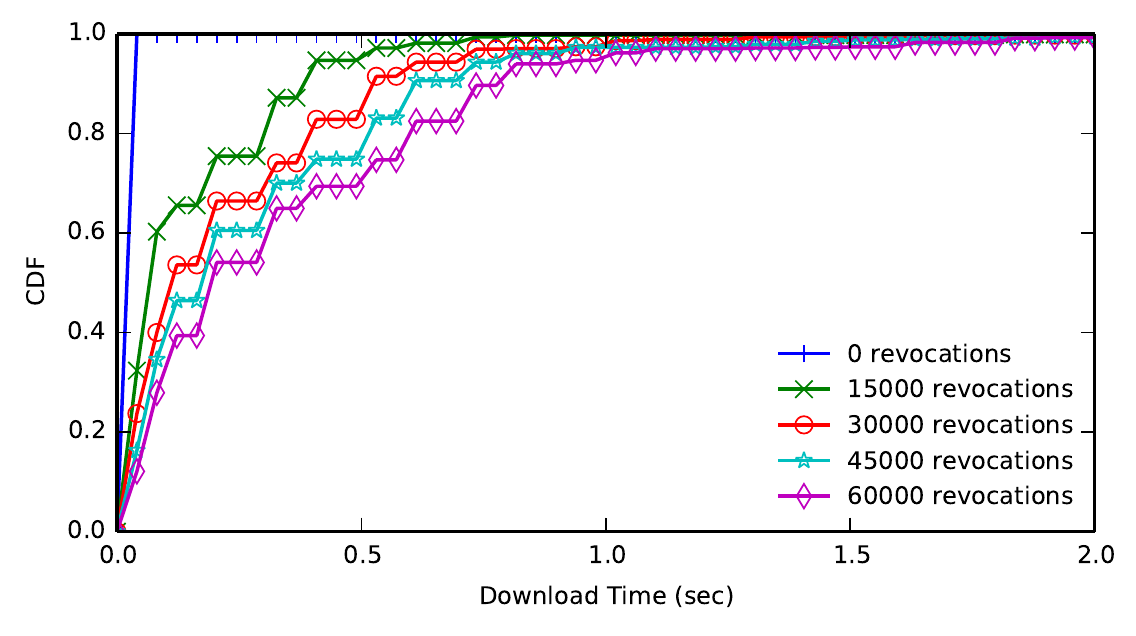}
    \vspace{-2pt}
    \caption{CDF of download times for five different revocation messages.}
    \label{fig:dissem-speed}
\end{figure}

For measurements, we create revocation messages of different sizes, and upload
them to Amazon CloudFront. More specifically, we test for five different
revocation messages: only a freshness statement (without revocations), and
revocation messages with 15K, 30K, 45K, and 60K revoked certificates. Then, from
80 Planetlab nodes from different geographical areas, we download the revocation
messages and measure the time it took to download the messages. We repeat this
experiment 10 times from each of the Planetlab nodes and for each of the five
revocation messages. The measured download times are shown as CDFs in
Fig.~\ref{fig:dissem-speed}.

Our measurements show that even for a large number of revocation messages (e.g.,
the purple line with diamond markers) and even for the worst-case scenario
(i.e., revocations are not cached), 90\% of the Planetlab nodes took less than
one second to download the revocations.

\subsection{Cost of Dissemination}\label{sec:eval:cost}

CDN operators charge content providers based on the amount of traffic that
travels in and out of their network. In our case, a CA is a content provider
that uploads the revocations and freshness statements and pays the CDN network
for the traffic that RAs generate to download the data. In this section, using
Amazon CloudFront as an example of CDN, we evaluate the price that a CA would
need to pay to the CDN operator.

For this evaluation, we need a time series of the number of certificates revoked
by a CA and the number and distribution of RAs deployed around the world. The
distribution of RAs is necessary because Amazon CloudFront has different pricing
rates for different locations~\cite{amz_link}.

For the former information, we use the largest CRL file that we could find (in
terms of number of distinct revocations)~\cite{crl_link}. It contains 339,557
revocation entries (almost 25\% of all the revocations we gathered) with a total
size of 7.5 MB.  This case is extreme, as we determine the average number of
revocations to be 5,440 per CRL.

There is no easy way to determine the number of RAs and their geographical
distribution. One possible approach is to estimate the number of RAs with the
number of IP addresses assigned to different parts of the world. However, such
an estimation may be inaccurate since NATs are widely deployed. Instead, we use
city population statistics; i.e., we estimate that the number of RAs is
proportional to the population size and consider geographical locality at the
granularity of cities. For the city population data, we use the
MaxMind~\cite{max_link} dataset, which reports the population size as well as
the latitude and longitude coordinates of the cities. The dataset has
information for 2.3 billion people from 47,980 cities. Then, we determine the
location of the edge servers that would serve the RAs in each city. Since we do
not have access to computers in each of the city, we use the closest Planetlab
node from each city to determine an edge server that would server an RA.

\begin{figure}
  \centering
    \includegraphics[width=0.8\linewidth]{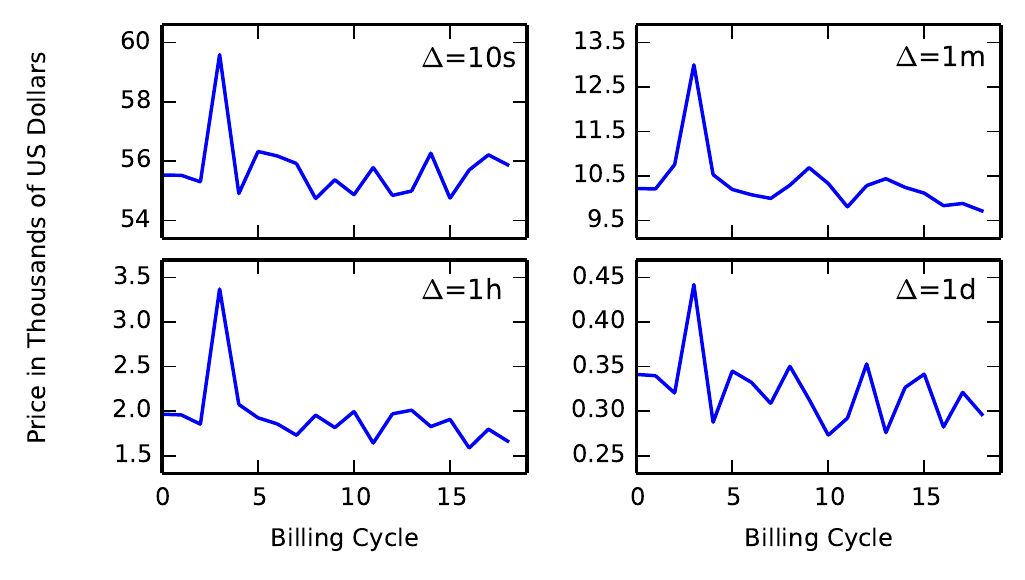}
    \caption{Monthly bills for a CA using a CDN for disseminating revocation
    lists (where there are 10 clients per RA).}
    \label{fig:rev-cost}
\end{figure}

Our cost estimates are conservative:
\begin{inparaenum}[\itshape 1)]
	\item we use the standard Amazon pricing, but when a CA negotiate with Amazon,
	  the pricing should be much lower,
	\item we assume that every RA serves only ten clients (thus there are 230
    million RAs in total),
	\item we analyze small values of $\Delta$, and
	\item the revocation list considered is the largest found.
\end{inparaenum}
In reality, the monthly operating cost for the CA should be lower than our
simulation, making CDN-based deployment even more attractive.

Fig.~\ref{fig:rev-cost} shows the monthly bill that the CA would have to pay
from 1 January 2014 to 1 August 2015, if they were to use \name to distribute
revocations. The simulation period includes the Heartbleed incident that
occurred in April 2014 to account for any catastrophic event that results in a
large number of revocations, and we show the results for four different values
of $\Delta$. We also estimate the average cost for the same CA, depending on how
many clients a single RA handles and the results are presented in
Tab.~\ref{tab:cost}. The values are given for small network (30 clients), large
network (250 clients), and a middle-size corporate network (1,000 clients).

\begin{table}
	\centering
	\footnotesize
	\begin{tabular}{@{}ccccc@{}}
		\toprule
		\centering \textbf{Clients per RA} & $\Delta=$\textit{10 sec} &
    $\Delta=$\textit{1 min} & $\Delta=$\textit{1 h} & $\Delta=$\textit{1 day} \\
		\midrule
		30 & 18.574 & 3.450 & 0.647 & 0.108\\
		250 & 2.229 & 0.414 & 0.078 & 0.013\\
		1,000 & 0.557 & 0.103 & 0.019 & 0.003\\
		\bottomrule
	\end{tabular}
	\caption{Average cost (in thousands of USD) in function of $\Delta$ and the
  number of clients per RA.}
	\label{tab:cost}
\end{table}

\subsection{Overheads}\label{sec:eval:overheads}

\myparagraph{Storage}
Distribution points, RAs, and CAs are the only parties that store revocations in
\name.  Neither clients nor servers need to store any revocation messages. This
property makes \name an ideal solution for hardware-limited clients (e.g., in
the Internet of Things) that are not able to store all complete and synchronized
revocation lists.  Moreover, through \name's design, the number of RAs should be
substantially smaller than the number of TLS clients and servers.

With the above-mentioned dataset and setting, we estimate the storage overhead
required by RAs to store the revocations. Conservatively, we assumed that all
revocations (even for expired certificates) are held by an RA. We also assumed
that dictionaries are implemented as described in \S\ref{sec:details:dict}. In
such a setting, the storage overhead is slightly above 4 MB and the memory
required to build and keep all dictionaries is 36 MB (for instance, for 10
million revocations this overhead is 30 MB and 260 MB accordingly).

\myparagraph{Communication}
\name introduces a negligible communication overhead to the TLS connection. The
only additional information sent during the TLS handshake is a proof of a
revocation status. The size of this proof is logarithmically proportional to the
number of revocations in a given dictionary, but it also depends on the
cryptographic primitives (in our case hashes are 20 bytes long). In our setting,
a revocation status (Eq.~\ref{eq:proof_root}) for an entry corresponding to the
largest CRL that we observed would be 500-900 bytes. This short message is sent
only during the connection establishment and then every $\Delta$.

\begin{figure}
  \centering
  \includegraphics[width=0.75\linewidth]{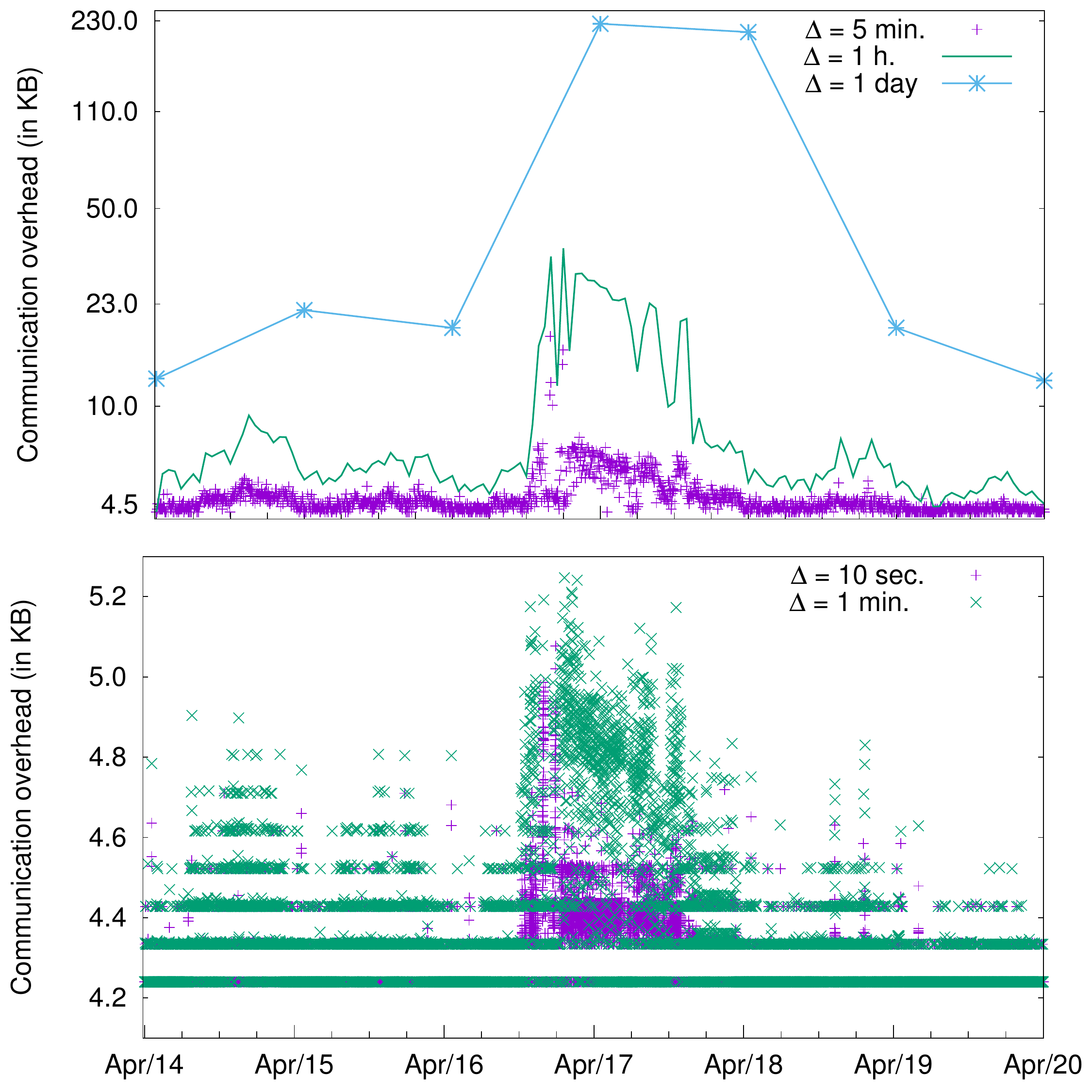}
  \vspace{-2pt}
  \caption{Communication overhead depending on a revocation rate and the
  $\Delta$ parameter. Note that the top y-axis is in logarithmic scale.}
  \label{fig:ra-bw}
\end{figure}

Communication overhead within the dissemination network is determined by
\begin{inparaenum}[\itshape 1)]
  \item the revocation rate (as RAs have to receive every revocation),
  \item the number of CAs (more precisely, the number of dictionaries), and
  \item the $\Delta$ parameter.
\end{inparaenum}
To estimate the required bandwidth, we analyze revocations issued during the
week in which the Heartbleed vulnerability was disclosed.  This period includes
standard and extremely high revocation rates. Conservatively, we assume the
number of dictionaries to be 254 (the number of CRLs we found), and we show
results for $\Delta$ equals 10 seconds, 1 minute, 5 minutes, 1 hour, and 1 day.
The obtained results are presented in Fig.~\ref{fig:ra-bw}, that depicts how
much data a single RA must download every $\Delta$.

As depicted, for a standard revocation rate, the required bandwidth is about 4
KB/$\Delta$. This overhead is mainly determined by the size of the freshness
statements and number of dictionaries (as every dictionary has to be refreshed
every $\Delta$). For the highest peak observed, the dissemination requires below
5 KB/$\Delta$ (for small $\Delta$s), around 25 KB for $\Delta$ equals 1 hour,
and about 230 KB when the RA is updated every day.

\myparagraph{Computation}
Using our implementation we evaluate the computational overhead introduced by
\name. Each test was executed 500 times on a machine with an Intel i7-4790
(3.60GHz) CPU, 16 GB of RAM, and running a 64-bit Linux.

First, we determine the time required to perform a dictionary update (which is
periodically conducted by CAs and RAs). For instance, to insert (see
Fig.~\ref{fig:auth_dict_api}) 1,000 new revocations, a CA needs to perform
computations, which take 3.88, 2.75, and 2.93 ms, on maximum, minimum, and
average, respectively. Similarly, for an RA to update a dictionary (see
Fig.~\ref{fig:auth_dict_api}) with 1,000 new revocations and then verify the
consistency of the dictionary and the authenticity of revocation issuance
message, it takes 5.87, 2.62, and 2.84 ms.

We also analyze the latency introduced to the TLS connection. This latency is
caused by computations executed by RAs and clients, and it is one of the most
important metrics for connection establishment, as it has a direct influence on
user reactions. The only additional latency in \name is introduced by DPI (RA),
proof construction (RA), and proof validation (client). We determine the
computational time required for these operations. The detailed results are
presented in Tab.~\ref{tab:comput}.

The RA needs to check whether each packet belongs to a TLS communication.
\begin{table}
	\centering
	\footnotesize
	\begin{tabular}{@{}lm{95pt}rrr@{}}
	\toprule
	\textbf{Entity} & \textbf{Operation} & \textbf{Max.}  & \textbf{Min.} &
  \textbf{Avg.} \\
	\midrule
	RA & TLS detection (DPI) & 14.07 & 1.90 & 2.93 \\
	RA & Certificates parsing (DPI)&57.94&15.97 &19.95 \\
	RA & Proof construction & 156.16 & 31.00 & 67.17 \\
	Client & Proof validation & 101.09 & 46.01 & 54.51 \\
	Client & Sig. and freshness valid. & 380.04 & 159.03 & 197.27 \\
	\bottomrule
	\end{tabular}
	\caption{Detailed processing time (in $\mu$s).}
	\label{tab:comput}
\end{table}
With our implementation, this verification takes on average 2.93 $\mu$s. RA must
parse every \texttt{ServerHello} message to obtain the server's certificate. The
computation time depends on the number of certificates in the chain (see
\S\ref{sec:discussion}), their length, used extensions, etc... For three
certificates in the chain (the most common number~\cite{ristic2011state}), such
parsing takes on average 19.95 $\mu$s. With the obtained certificate, the RA
checks its revocation status, and returns the corresponding proof. This
operation takes 67.17 $\mu$s on average. Then, the revocation status obtained by
the client must be verified. First, the client verifies whether the proof
matches the obtained certificate and root (from the revocation status message).
Then, the client verifies the signature of the signed root and corresponding
freshness statement. These operations take on average 54.51 and 197.27 $\mu$s,
respectively. The overall overhead introduced on the client side is about 250
$\mu$s. In a contrast to these results, a study showed that even an optimized
TLS handshake takes about 30 ms (with network latency)~\cite{stark2012case}.
Therefore, the overhead introduced by \name to TLS connection establishment is
less than 1\%.

As the result, an RA can process more than 340,000 non-TLS packets per second
and more than 50,000 \name-supported TLS handshakes per second, on average.
Clients can validate almost 4,000 revocation statuses per second. Thus, the
overhead introduced by \name should not be noticed by clients.

\subsection{Comparison}\label{sec:eval:comparison}
\begin{table}[b]
	\centering
	\footnotesize
	\begin{tabular}{@{}lrrrrr@{}}
		\toprule
		& \textbf{Storage} & \textbf{Storage} & \textbf{Conn.} &\textbf{Conn.} & \textbf{Violated} \\
		\textbf{Method} & \textbf{(global)} & \textbf{(client)} & \textbf{(global)} &\textbf{(client)} & \textbf{properties} \\
		\midrule
		CRL           & $n_{rev} \times (n_{cl}+1)$    & $n_{rev}$ & $n_{cl} \times n_{ca}$ & $n_{ca}$  & I, P, E, T \\
		CRLSet$^a$    & $n_{rev} \times (n_{cl}+1)$    & $n_{rev}$ & $n_{cl}$               & 1         & I, E, T \\
		OCSP          & $n_{rev}$                      & 0         & $n_{cl} \times n_s$    & $n_s$     & I, P, E, T \\
		OCSP$^b$      & $n_{rev}+n_s$                  & 0         & $n_s$                  & 0         & I, S, T \\
		Log$^c$       & $n_{rev}$                      & 0         & $n_{cl} \times n_s$    & $n_s$     & I, P, E\\
		Log$^d$       & $n_{rev}$                      & 0         & $n_s$                  & 0         & I, S \\
		RevCast$^e$   & $n_{rev} \times (n_{cl}+1)$    & $n_{rev}$ & $n_{cl}$               & $n_{rev}$ & E, T \\
		\name         & $n_{rev} \times (n_{ra}+1)$    & 0         & $n_{ca}$               & 0         & - \\
		\bottomrule
	\end{tabular}
	\caption{Comparison of revocation mechanisms in terms of properties and overhead.
	$n_s,n_{ca},n_{ra},n_{cl},n_{rev}$: number of servers, CAs, RAs, clients, and
	revocations, respectively. $n_{ca}\ll n_{ra}<n_s\ll n_{cl}$.\\
	I: near-instant revocation  P: privacy  E: efficiency and scalability
	T: transparency and accountability  S: server changes not required.
	$^a$CRLSets contain a limited number of revocations.
	$^b$OCSP Stapling.
	$^c$Client-driven approaches.
	$^d$Server-driven approaches.
	$^e$RevCast uses radio broadcast for dissemination.}
	\label{tab:comparison}
\end{table}

We compare \name with competing schemes (see \S\ref{sec:pre}) and present the
results in Tab.~\ref{tab:comparison}. We consider the storage overhead, the
number of connections required by the scheme, and the achieved properties. For
this comparison, we assume that the given revocation scheme is fully deployed.
Then, we show the storage and the number of connections required to achieve a
state in which an arbitrary client is able to establish a secure connection with
an arbitrary server (i.e., the client must learn about the revocation status of
the server's certificate). Log-based approaches are presented in their two
deployment models (see \S\ref{sec:related}). Global values show how the complete
revocation list is replicated, and how many connections are required in total.
Per-client values show how many entries a client must store, and how many
messages it must obtain. The comparison demonstrates the scalability of \name.
Clients and servers neither store nor request for any messages. OCSP Stapling
and log-based approaches (with server-driven deployment) have good performance
results, but their deployment depends on the server's configuration, and they
lack other desired properties.

\subsection{Deployability}\label{sec:eval:deploy}
\name has many advantages from a deployability point of view. An RA must provide
a set of simple functionalities and can be cheaply implemented as a stand-alone
device or as a part of another network device. Through the different deployment
models that we present in \S\ref{sec:deploy}, we show that \name's deployment
can combine both, efficiency and security, and that \name can be incrementally
deployed. One RA can protect all TLS connections from/to a whole network.
Therefore, a mass deployment is possible with a low operational effort. The
deployment can be driven by a set of individual parties such as LAN operators,
ISPs, or server farms operators. \name is also backward-compatible as RAs are
completely non-invasive for non-supported clients and protocols other than TLS.

%% file: discussion.tex
\myparagraph{Bootstrapping CAs into \name}\label{sec:deploy:incre}
Current CAs that wish to start \name deployment could publish a signed manifest
at pre-defined locations (e.g., \url{/RITM.json}). Such a manifest would contain
the CDN address of a dictionary.  Then, every RA would periodically check (e.g.,
once per week) whether a CA started \name deployment.  Information about such a
CA should be provided to the clients as well, and this could be realized through
similar periodic checks or through software update (as the manifest is short).

\myparagraph{Local $\Delta$ parameter} To simplify the presentation, we assumed
that the $\Delta$ parameter was globally agreed on. However, to relax this
assumption, each CA could express its own $\Delta$ parameter in a dedicated
field of its certificate (or in a manifest as proposed above). Then, clients and
RAs (who need CA certificates to validate server certificates or/and revocation
statuses anyway) would know the correct $\Delta$ value, and operate accordingly.

\myparagraph{Ever-growing dictionaries} We also assumed that each CA has a
single append-only dictionary. Consequently, revocation entries (even for
expired certificates) cannot be removed.  However, this assumption can be
relaxed as well. For instance, a CA can split all revocations into a few
dictionaries.  A similar practice is popular in maintaining CRLs (to reduce
bandwidth overhead), and although it is not necessary for \name, this can limit
storage used on RAs.  More precisely, a CA could maintain a few dictionaries at
the same time, and every dictionary would be dedicated to all the certificates
that expire before a given time.  As ``certificates issued after 1 April 2015
must have a validity period no greater than 39 months'' (according to the CA/B
Forum~\cite{CABForum}), RAs could then regularly delete the dictionaries that
correspond to expired certificates.

\myparagraph{Certificate chains} We presented \name in a setting where the
validity proof is provided only for a server certificate. In practice, a chain
of certificates is used. Although many revocation schemes ignore the revocation
checking of CA certificates, \name can be extended to support this feature.
Instead of returning a single absence proof, an RA would return a proof for each
certificate of the chain.  The introduced overhead should not increase
significantly, as certificate chains are usually short~\cite{ristic2011state},
and as the proof construction is an efficient operation (see
\S\ref{sec:eval:overheads}).

\myparagraph{Multiple RAs} As RAs can be installed independently, it may happen
that a single \name-supported connection traverses multiple RAs. To maintain a
constant overhead in such a situation, we require that an RA adds a revocation
status only when it is missing, and replaces a revocation status only if its own
version of the dictionary is more recent. Whenever an RA sees a revocation
message sent by another RA, it can verify whether their view of the dictionary
is consistent, by simply comparing the two versions of the signed roots.

\myparagraph{RA-to-client communication}
One challenge in implementing \name is to reliably transfer a revocation status
from an RA to a client along with a \texttt{ServerHello} message. Fortunately,
middleboxes have been studied for years, and we can benefit from previous works
and well-known techniques. We distinguish three possible methods:
\begin{enumerate}[leftmargin=*]
  \item The status is piggybacked on the native TLS message. Since, the payload
    of the TCP packet (i.e., the packet that carries the altered TLS message)
    must be extended, the RA must adjust the sequence numbers of the TCP
    session. Then, the RA must also indicate (e.g., through a dedicated TLS
    \texttt{Content Type}) that client should handle the TLS message
    differently. Otherwise TLS would refuse the connection, as the TLS handshake
    was modified.
  \item The client opens a dedicated port, to which RA sends \name messages.
    Though, such a design is easy to realize, it does not allow to contact
    clients placed behind a NAT.
  \item Instead of using a dedicated port, the port that the client is using for
    the TLS communication is used to send the revocation status. This approach
    solves the NAT problem, and does not influence the TLS protocol directly,
    but RAs still have to adjust the TCP state for supported sessions.
\end{enumerate}

\myparagraph{Future work} We will investigate the possible interactions between
\name and novel PKI enhancements~\cite{PoliCert,PKISN}. We will also analyze
some aspects of the incremental deployment of technologies such as \name. Other
topics of interest include a study on ensuring the consistency between RAs and
clients, and applying novel deployment models~\cite{sherry2012}, where \name can
be implemented as a service.

%% file: conclusions.tex
In this paper, we present \name, a framework that aims to address the revocation
problem through distinctive properties of middleboxes and content delivery
networks. \name takes advantage of an existing infrastructure to disseminate
revocation messages, and moves certificate revocation lists to the middleboxes.
Such a combination results in a scheme that satisfies all the requirements we
identified. In particular, we believe that \name is the first scheme that can
provide near-instantaneous revocation in a real-world deployment. Among other
advantages, \name prepares us for future demands originating from, e.g.,
pervasive encryption or the Internet of Things, in which countless
hardware-constrained devices would need to communicate securely.

%% file: paper.bbl
\begin{thebibliography}{10}

\bibitem{amz_link}
{Amazon CloudFront}.
\newblock \url{https://aws.amazon.com/cloudfront}.

\bibitem{crl_link}
{CAcert's CRL}.
\newblock \url{http://crl.cacert.org/revoke.crl}.

\bibitem{hb_link}
The {Heartbleed} bug.
\newblock \url{http://heartbleed.com}.

\bibitem{max_link}
{MaxMind}.
\newblock \url{https://www.maxmind.com}.

\bibitem{mozilla_rev}
Mozilla's revocation plan.
\newblock \url{https://wiki.mozilla.org/CA:RevocationPlan}.

\bibitem{CABForum}
Baseline requirements for the issuance and management of publicly-trusted
  certificates.
\newblock
  \url{https://cabforum.org/wp-content/uploads/Baseline_Requirements_V1_1_6.pdf},
  2014.
\newblock CA/Browser Forum.

\bibitem{chrome_crl}
An evaluation of the effectiveness of {Chrome}'s {CRLSets}.
\newblock \url{https://www.grc.com/revocation/crlsets.htm}, 2014.

\bibitem{letsencrypt}
{Let's Encrypt}.
\newblock \url{https://letsencrypt.org/}, 2014.

\bibitem{isc_crl}
{SSL} {CRL} activity.
\newblock \url{https://isc.sans.edu/crls.html}, 2015.
\newblock {SANS Internet Storm Center}.

\bibitem{al2011overclocking}
M.~Al-Fares, K.~Elmeleegy, B.~Reed, and I.~Gashinsky.
\newblock Overclocking the {Yahoo!} {CDN} for faster web page loads.
\newblock In {\em ACM IMC}, 2011.

\bibitem{Basin:2014:AAR:2660267.2660298}
D.~Basin, C.~Cremers, T.~H.-J. Kim, A.~Perrig, R.~Sasse, and P.~Szalachowski.
\newblock {ARPKI}: Attack resilient public-key infrastructure.
\newblock In {\em ACM CCS}, 2014.

\bibitem{bernstein2012high}
D.~J. Bernstein, N.~Duif, T.~Lange, P.~Schwabe, and B.-Y. Yang.
\newblock High-speed high-security signatures.
\newblock {\em Journal of Cryptographic Engineering}, 2012.

\bibitem{gossip2015}
L.~Chuat, P.~Szalachowski, A.~Perrig, B.~Laurie, and E.~Messeri.
\newblock Efficient gossip protocols for verifying the consistency of
  certificate logs.
\newblock In {\em IEEE CNS}, 2015.

\bibitem{chun2003planetlab}
B.~Chun, D.~Culler, T.~Roscoe, A.~Bavier, L.~Peterson, M.~Wawrzoniak, and
  M.~Bowman.
\newblock {PlanetLab}: an overlay testbed for broad-coverage services.
\newblock {\em ACM SIGCOMM Computer Communication Review}, 2003.

\bibitem{rfc5280}
D.~Cooper, S.~Santesson, S.~Farrell, S.~Boeyen, R.~Housley, and W.~Polk.
\newblock Internet {X.509} public key infrastructure certificate and
  certificate revocation list ({CRL}) profile.
\newblock RFC 5280, May 2008.

\bibitem{delignat2015network}
A.~Delignat-Lavaud and K.~Bhargavan.
\newblock Network-based origin confusion attacks against {HTTPS} virtual
  hosting.
\newblock In {\em WWW}, 2015.

\bibitem{rfc5246}
T.~Dierks and E.~Rescorla.
\newblock The transport layer security ({TLS}) protocol version 1.2.
\newblock RFC 5246, August 2008.

\bibitem{Droms:2001:ADM:RFC3118}
R.~Droms and W.~Arbaugh.
\newblock Authentication for {DHCP} messages.
\newblock RFC 3118, June 2001.

\bibitem{durumeric2014matter}
Z.~Durumeric, J.~Kasten, D.~Adrian, J.~A. Halderman, M.~Bailey, F.~Li,
  N.~Weaver, J.~Amann, J.~Beekman, and M.~Payer.
\newblock The matter of {Heartbleed}.
\newblock In {\em ACM IMC}, 2014.

\bibitem{rfc7469}
C.~Evans, C.~Palmer, and R.~Sleevi.
\newblock Public key pinning extension for {HTTP}.
\newblock RFC 7469, April 2015.

\bibitem{gruschka2014analysis}
N.~Gruschka, L.~L. Iacono, and C.~Sorge.
\newblock Analysis of the current state in website certificate validation.
\newblock {\em Security and Communication Networks}, 2014.

\bibitem{gutmann2004build}
P.~Gutmann.
\newblock How to build a {PKI} that works.
\newblock In {\em PKI R\&D Workshop}, 2004.

\bibitem{rfc6797}
J.~Hodges, C.~Jackson, and A.~Barth.
\newblock {HTTP} strict transport security ({HSTS}).
\newblock RFC 6797, November 2012.

\bibitem{KimHuaPerJacGli13}
T.~H.-J. Kim, L.-S. Huang, A.~Perrig, C.~Jackson, and V.~Gligor.
\newblock Accountable key infrastructure ({AKI}): A proposal for a public-key
  validation infrastructure.
\newblock In {\em WWW}, 2013.

\bibitem{langley2012revocation}
A.~Langley.
\newblock Revocation checking and {Chrome}'s {CRL}.
\newblock \url{https://www.imperialviolet.org/2012/02/05/crlsets.html}, 2012.

\bibitem{laurie2012revocation}
B.~Laurie and E.~Kasper.
\newblock Revocation transparency, 2012.

\bibitem{liu-2015-revocation}
Y.~Liu, W.~Tome, L.~Zhang, D.~Choffnes, D.~Levin, B.~Maggs, A.~Mislove,
  A.~Schulman, and C.~Wilson.
\newblock An end-to-end measurement of certificate revocation in the web's
  {PKI}.
\newblock In {\em ACM IMC}, 2015.

\bibitem{marlinspike2009more}
M.~Marlinspike.
\newblock More tricks for defeating {SSL} in practice.
\newblock {\em Black Hat USA}, 2009.

\bibitem{naor2000certificate}
M.~Naor and K.~Nissim.
\newblock Certificate revocation and certificate update.
\newblock {\em IEEE Journal on Selected Areas in Communications}, 2000.

\bibitem{pettersen2013transport}
Y.~Pettersen.
\newblock The transport layer security ({TLS}) multiple certificate status
  request extension.
\newblock RFC 6961, June 2013.

\bibitem{hidden_cost}
M.~Prince.
\newblock The hidden costs of {Heartbleed}.
\newblock \url{https://blog.cloudflare.com/the-hard-costs-of-heartbleed/},
  2014.

\bibitem{universal_ssl}
M.~Prince.
\newblock {Universal SSL}.
\newblock \url{http://blog.cloudflare.com/introducing-universal-ssl/}, 2014.

\bibitem{Qazi2013}
Z.~A. Qazi, C.-C. Tu, L.~Chiang, R.~Miao, V.~Sekar, and M.~Yu.
\newblock {SIMPLE-fying} middlebox policy enforcement using {SDN}.
\newblock In {\em ACM SIGCOMM}, 2013.

\bibitem{rescorla2016transport}
E.~Rescorla.
\newblock The transport layer security (tls) protocol version 1.3.
\newblock 2016.

\bibitem{ristic2011state}
I.~Ristic.
\newblock State of {SSL}.
\newblock {\em InfoSec World}, 2011.

\bibitem{rivest1998can}
R.~L. Rivest.
\newblock Can we eliminate certificate revocation lists?
\newblock In {\em Financial Cryptography}, 1998.

\bibitem{rfc5077}
J.~Salowey, H.~Zhou, P.~Eronen, and H.~Tschofenig.
\newblock Transport layer security ({TLS}) session resumption without
  server-side state.
\newblock RFC 5077, January 2008.

\bibitem{santesson2013x}
S.~Santesson, M.~Myers, R.~Ankney, A.~Malpani, S.~Galperin, and C.~Adams.
\newblock {X.509} {Internet} public key infrastructure online certificate
  status protocol -- {OCSP}.
\newblock RFC 6960, June 2013.

\bibitem{schulman2014revcast}
A.~Schulman, D.~Levin, and N.~Spring.
\newblock {RevCast}: Fast, private certificate revocation over {FM} radio.
\newblock In {\em ACM CCS}, 2014.

\bibitem{sherry2012}
J.~Sherry, S.~Hasan, C.~Scott, A.~Krishnamurthy, S.~Ratnasamy, and V.~Sekar.
\newblock Making middleboxes someone else's problem: Network processing as a
  cloud service.
\newblock In {\em ACM SIGCOMM}, 2012.

\bibitem{sslpulse2015}
{SSL Labs}.
\newblock {SSL Pulse}.
\newblock \url{https://www.trustworthyinternet.org/ssl-pulse/}, 2012.

\bibitem{stark2012case}
E.~Stark, L.-S. Huang, D.~Israni, C.~Jackson, and D.~Boneh.
\newblock The case for prefetching and prevalidating {TLS} server certificates.
\newblock In {\em NDSS}, 2012.

\bibitem{PKISN}
P.~Szalachowski, L.~Chuat, and A.~Perrig.
\newblock {PKI} safety net ({PKISN}): Addressing the too-big-to-be-revoked
  problem of the {TLS} ecosystem.
\newblock In {\em IEEE EuroS\&P}, 2016.

\bibitem{PoliCert}
P.~Szalachowski, S.~Matsumoto, and A.~Perrig.
\newblock {PoliCert}: Secure and flexible {TLS} certificate management.
\newblock In {\em ACM CCS}, 2014.

\bibitem{topalovic2012towards}
E.~Topalovic, B.~Saeta, L.-S. Huang, C.~Jackson, and D.~Boneh.
\newblock Towards short-lived certificates.
\newblock {\em Web 2.0 Security and Privacy}, 2012.

\bibitem{Zhang:2014:ASC:2663716.2663758}
L.~Zhang, D.~Choffnes, D.~Levin, T.~Dumitras, A.~Mislove, A.~Schulman, and
  C.~Wilson.
\newblock Analysis of {SSL} certificate reissues and revocations in the wake of
  {Heartbleed}.
\newblock In {\em ACM IMC}, 2014.

\end{thebibliography}
